\documentclass[twocolumn,secnumarabic,amssymb, nobibnotes, aps, prl]{revtex4-2}
\usepackage{amsfonts,amssymb,amsmath} %%%add
\usepackage{color} %%%add
\usepackage{graphicx} %%%add
\usepackage{epstopdf,mathrsfs} %%%add
\usepackage[colorlinks,linktocpage,linkcolor=blue]{hyperref}
\begin{document}

%Langevin picture of subdiffusion in different external force field

\title{Theory of relaxation dynamics for anomalous diffusion processes in harmonic potential}

\author{Xudong Wang}
\author{Yao Chen}
\author{Weihua Deng}

\affiliation{School of Mathematics and Statistics, Gansu Key Laboratory
of Applied Mathematics and Complex Systems, Lanzhou University, Lanzhou 730000,
P.R. China}

\begin{abstract}
Optical tweezers setup is often used to probe the motion of individual tracer particle, which promotes the study of relaxation dynamics of a generic process confined in a harmonic potential.
We uncover the dependence of ensemble- and time-averaged mean square displacements of confined processes on the velocity correlation function $C(t,t+\tau)$ of the original process.
With two different scaling forms of $C(t,t+\tau)$ for small $\tau$ and large $\tau$, the stationary value and the relaxation behaviors can be obtained immediately.
The gotten results are valid for a large amount of anomalous diffusion processes, including fractional Brownian motion, scaled Brownian motion, and the multi-scale L\'{e}vy walk with different exponents of running time distribution.
\end{abstract}

\pacs{}

\maketitle

%\section{Introduction}
The anomalous diffusion phenomena are ubiquitous in the natural world, especially in numerous microscopic systems. It is in general characterized by the nonlinear evolution in time of the ensemble-averaged mean squared displacement (EAMSD), i.e., $\langle x^2(t)\rangle \propto t^\alpha$ with $\alpha\neq1$ \cite{Bouchaud:1992,MetzlerKlafter:2000}. Apart from EAMSD, another important quantity to detect particle-to-particle diffusion properties is time-averaged mean squared displacement (TAMSD), defined as  \cite{MetzlerJeonCherstvyBarkai:2014}
\begin{equation}\label{TADef}
\overline{\delta^2(\Delta)}=\frac{1}{T-\Delta}\int_0^{T-\Delta} [x(t+\Delta)-x(t)]^2dt,
\end{equation}
which can be obtained by analyzing the time series of single trajectory in experiments. Here $\Delta$ is the lag time, assumed to be much shorter than the total measurement time $T$ to obtain good statistical properties.

The single particle tracking techniques have been widely employed to study diffusion of particles in living cell \cite{GoldingCox:2006,WeberSpakowitzTheriot:2010,BronsteinIsraelKeptenMaiTalBarkaiGarini:2009}.
A process is called ergodic if the two MSDs are equal, i.e., $\overline{\delta^2(t)}=\langle x^2(t)\rangle$ as $T\rightarrow\infty$, such as Brownian motion. An interesting phenomenon accompanied by Brownian motion is the discrepancy between EAMSD and TAMSD, if it is confined in a harmonic potential \cite{JeonMetzler:2012}. Coincidentally, the ergodic fractional Brownian motion (FBM) \cite{DengBarkai:2009} and fractional Langevin equation also present the same phenomenon---the TAMSD converges to twice of the EAMSD as $T$ and $\Delta$ tend to infinity %$T,\Delta\rightarrow\infty$
 \cite{JeonMetzler:2012}. By contrast, different diffusion properties can be observed for a confined non-ergodic process, such as the continuous-time random walk (CTRW) with diverging characteristic waiting times. Under confinement, the EAMSD of this subdiffusive process converges to its stationary value $\langle x^2\rangle_{\textrm{th}}$ while the TAMSD behaves as $\Delta^{1-\alpha}$ \cite{BurovMetzlerBarkai:2010}.

From the point of view of uncovering the mechanism, it seems necessary to study the stationary value as well as the relaxation dynamics of the generic processes confined in harmonic potential, especially those displaying anomalous diffusion behavior. %Moreover, it is an important physical property for a stochastic process how it responses to an external force or spatial confinement.
Moreover, how to respond to an external force or spatial confinement is an important physical property for a stochastic process.
On the other hand, from the point of view of experimental observation, the tracer particles are recorded in an optical tweezers setup, which exerts a restoring Hookean force on the particle \cite{Jeon.etal:2011}.

Among the numerous models describing anomalous diffusion phenomenon, one special representative is L\'{e}vy walk \cite{ZaburdaevDenisovKlafter:2015}. L\'{e}vy walk dynamics theoretically show the enhanced transport phenomena in many systems, such as diffusion in Josephson junctions \cite{GeiselNierwetbergZacherl:1985} and diffusion of atoms in optical lattices \cite{MarksteinerEllingerZoller:1996,BarkaiAghionKessler:2014}. The exponent $\alpha$ of the running time distribution decides the diffusion behavior of L\'{e}vy walk.
It says that L\'{e}vy walk is ultraweak nonergodic \cite{FroembergBarkai:2013,*GodecMetzler:2013} for $\alpha$ in $(0,1)$ and $(1,2)$.
Then a question naturally arises: What about the ergodic behavior of the confined L\'{e}vy walk with different $\alpha$? In this letter, we derive the direct relationship between the relaxation dynamics of any confined process, not just limited to L\'{e}vy walk, and the scaling form of the corresponding free particle's velocity correlation function (VCF) $C(t,t+\tau)$, which is usually nonstationary for a nonergodic process.

The essential characteristics of a stochastic process are usually embodied by its correlation function. A nonstationary process is accompanied with an aging correlation function. A wide class of physical systems exhibit a kind of correlation function scaling as $C(t,t+\tau)\simeq t^\gamma\phi(\tau/t)$ for large time $t$ and lag time $\tau$.
Based on this scaling form, the generalized Green-Kubo formula \cite{DechantLutzKesslerBarkai:2014,*MeyerBarkaiKantz:2017} has been built, which reveals the dependence of EAMSD and TAMSD on the correlation function. In addition,
the relationship between the power spectrum and the correlation functions has also been formulated, which is named as aging Wiener-Khinchin theorem \cite{LeibovichBarkai:2015,*DechantLutz:2015}.

However, the situation becomes much different for a confined system. %Now the asymptotic behavior of correlation function with small $\tau$ plays a nonnegligible role.
The asymptotic behavior of the correlation function with small $\tau$ can no longer be omitted.
It is found that the TAMSD depends on both of the cases of large and small $\tau$ while the EAMSD only depends on the small one. Especially, the VCF of a nonstationary process might exhibit multi-scale forms for different range of lag time $\tau$, such as L\'{e}vy walk. Combining different scaling forms provides a comprehensive characterization of a stochastic process.

%\section{An overdamped Langevin equation with harmonic potential}
\emph{The generic results.}--- %The overdamped Langevin equation is an effective model to describe some process $x(t)$ confined in a harmonic potential $U(x)=kx^2/2$, as
The overdamped Langevin equation can effectively describe some process $x(t)$ confined in a harmonic potential $U(x)=kx^2/2$, which has the simple formulation
\begin{equation}\label{model1}
  \dot{x}(t)= -kx(t) + \xi(t).
\end{equation}
The random force $\xi(t)$ here can be any kind of process concerned. Equation (\ref{model1}) models Brownian motion or FBM in a harmonic potential if $\xi(t)$ is Gaussian white noise or fraction Gaussian noise.
 %It is Gaussian white noise or fraction Gaussian noise to model Brownian motion or FBM in a harmonic potential.
 The process $x(t)$ in \eqref{model1} can be analytically solved through the technique of Laplace transform as $x(t)=\int_0^t dt'e^{-k(t-t')}\xi(t')$, from which the position correlation function can be directly obtained as
\begin{equation}\label{CorrX}
\begin{split}
&\langle x(t_1)x(t_2)\rangle
=\int_0^{t_1}\!\! dt_1'\int_0^{t_2}\!\! dt_2'e^{-k(t_1-t_1')}e^{-k(t_2-t_2')}C(t_1',t_2'),
\end{split}
\end{equation}
where $C(t_1,t_2)=\langle \xi(t_1)\xi(t_2)\rangle$.
Equation \eqref{CorrX} shows that the correlation function of the random force $\xi(t)$ plays a crucial role in the moments of $x(t)$.  Especially, the EAMSD $\langle x^2(t)\rangle$ is determined by the asymptotic form of $C(t,t+\tau)$ for large $t$ and small $\tau$, i.e.,
\begin{equation}\label{Asym1}
  C(t,t+\tau) \simeq f_1(\tau)+ t^\gamma f_2(\tau),
\end{equation}
where the exponent $\gamma$ is assumed to be negative to avoid the growth of the correlation. A stationary process only possesses the first term. For complex non-stationary process, the correlation function explicitly depends on time $t$, and the second term decaying at a power law rate becomes nonnegligible.
With some calculations in Supplemental Material \cite{SM}, we have
\begin{equation}\label{EAMSD}
   \langle x^2(t)\rangle \simeq  \hat{f}_1(k)/k+t^\gamma\hat{f}_2(k)/k,
\end{equation}
where the terms decaying exponentially have been omitted, and $\hat{f}(k)=\int_0^\infty dt e^{-kt}f(t)$ is the Laplace transform of $f(t)$.

Equation \eqref{EAMSD} implies that the EAMSD converges to the stationary value $\langle x^2\rangle_{\textrm{th}}=\hat{f}_1(k)/k$ at the power law rate $t^\gamma$. The one-to-one correspondence between the correlation function $C(t,t+\tau)$ and the EAMSD  shows that the $t$-independent term in \eqref{Asym1} determines the stationary value $\langle x^2\rangle_{\textrm{th}}$ while the $t$-dependent term the relaxation behavior. Thus, the EAMSD converges to the thermal value  $\langle x^2\rangle_{\textrm{th}}$ exponentially with a single characteristic time scale $1/k$ for a stationary process in a harmonic potential. In contrast, for the nonstationary process with an aging correlation function $C(t,t+\tau)$, the corresponding confined EAMSD relaxes algebraically and the relaxation is usually very slow for the process presenting a significant aging phenomenon with $-1<\gamma<0$.
The stationary thermal value is mainly contributed by the small $\tau$ correlations as $t\rightarrow\infty$.
Sometimes, the first term in \eqref{Asym1} vanishes for the process without an inherent correlation as $t\rightarrow\infty$, such as scaled Brownian motion (SBM) \cite{ThielSokolov:2014,JeonChechkinMetzler:2014} with the correlation function of noise scaling as $t^\gamma\delta(\tau)$. In this case, the random force is too weak so that the thermal value becomes void.

As for the TAMSD defined in \eqref{TADef}, it can also be evaluated through the scaling form of correlation function $C(t,t+\tau)$. Here, we only consider the large $\Delta$ behavior of TAMSD, since the small asymptotic behavior will be the same as the free-force case. As usual, the scaling form of $C(t,t+\Delta)$ is assumed as, for large $t$ and $\Delta$,
\begin{equation}\label{Asym2}
  C(t,t+\Delta)\simeq t^\beta \phi\left(\frac{\Delta}{t}\right).
\end{equation}
When evaluating the TAMSD, we find that
\begin{equation}
  \langle x(t)x(t+\Delta)\rangle \simeq \frac{1}{k^2}C(t,t+\Delta),
\end{equation}
after omitting the terms decaying exponentially for large $t$ and $\Delta$. As the integrand of TAMSD with $\Delta\ll T$, the asymptotic form $\Delta\ll t$ of $C(t,t+\Delta)$ plays a dominating role. Assume that the scaling function $\phi$  has the asymptotic expansion as,
\begin{equation}\label{phix}
  \phi(x)\simeq c_1x^{\nu_1}+c_2x^{\nu_2}+\cdots,  \qquad  x\rightarrow0,
\end{equation}
where the exponents might be both positive or negative satisfying $\beta\leq\nu_1<\nu_2$. The lower bound ensures that the correlation \eqref{Asym2} will not increase with the growth of the time $t$. With some technical evaluations in Supplemental Material \cite{SM}, we obtain
\begin{equation}\label{TAMSD}
\begin{split}
    &\langle \overline{\delta^2(\Delta)} \rangle \simeq \frac{2\hat{f}_1(k)}{k}+\frac{\hat{f}_2(k)}{k} \frac{2T^\gamma}{\gamma+1}   \\
   &~~~ - \frac{2}{k^2} \left[\frac{c_1}{1+\beta-\nu_1}\frac{\Delta^{\nu_1}}{T^{\nu_1-\beta}}+\frac{c_2}{1+\beta-\nu_2}\frac{\Delta^{\nu_2}}{T^{\nu_2-\beta}}\right].
\end{split}
\end{equation}
The first two terms come from the EAMSD $\langle x^2(t)\rangle$ in \eqref{EAMSD} determined by the behavior of small $\tau$ in \eqref{Asym1} while the latter two terms depend on the behavior of large $\tau$ in \eqref{Asym2}. Especially, the latter two terms vanish if the correlation function $C(t,t+\tau)$ is a $\delta$-function, such as Brownian motion and SBM.
The third term is a leading term compared with the fourth one. We present both of the last two terms in case of the leading one becoming a constant when $\nu_1=\beta=0$.

Similarly to the derivation of \eqref{EAMSD}, we have ignored the terms decaying exponentially in \eqref{TAMSD}. Thus, the TAMSD relaxes to a stationary value exponentially only if the latter three terms vanish, i.e., some stationary process together with its correlation function of noise being a $\delta$-function or exponentially decaying, e.g., Brownian motion. In contrast to that, the TAMSD might relax to a saturated value algebraically for the general processes presenting anomalous diffusion.
The saturated value is usually $2\hat{f}(k)/k$, twice the stationary value $\langle x^2\rangle_{\textrm{th}}$ of the EAMSD, provided that the third term is not a constant. Otherwise, the saturated value will be much different, and even vanish for a scale-free process, such as the subdiffusive CTRW \cite{BurovMetzlerBarkai:2010}.

%\section{Examples}
\emph{FBM and SBM.}--- As a generalization of Brownian motion, both FBM and SBM inherit the character of Gaussian-shaped probability density function, which can solve the same Fokker-Planck equation with time-dependent coefficient. The process describing how they respond to a harmonic potential follows the Langevin equation \eqref{model1} with the random force satisfying
\begin{equation}\label{FBM}
  \langle \xi^\textrm{FBM}(t_1) \xi^\textrm{FBM}(t_2) \rangle \simeq \alpha K (\alpha-1)|t_1-t_2|^{\alpha-2}
\end{equation}
and
\begin{equation}\label{SBM}
  \langle \xi^\textrm{SBM}(t_1) \xi^\textrm{SBM}(t_2) \rangle \simeq 2\alpha K t_1^{\alpha-1}\delta(t_1-t_2),
\end{equation}
respectively. Though their correlation functions are different, for the free-force case, the EAMSDs are both $2Kt^\alpha$, displaying superdiffusion for $\alpha>1$ and subdiffusion for $\alpha<1$.

%FBM has stationary and correlated increments while SBM is nonstationary but Markovian. 

The increment of FBM is stationary and correlated while that of SBM is nonstationary and independent.
These contrary characteristics are related to the two typical single-scale correlation functions in \eqref{FBM} and \eqref{SBM}.
By comparing them with the scaling form of $C(t,t+\tau)$ in \eqref{Asym1} and \eqref{Asym2}, nonoverlapping parts can be found between them, i.e., $f_2(\tau)=0$ for FBM and $f_1(\tau)=\phi(x)=0$ for SBM. Therefore, the MSDs for them confined in a harmonic potential will be completely different.

The EAMSD of FBM in harmonic potential tends to the thermal value $\langle x^2\rangle_\textrm{th}=K\Gamma(\alpha+1)/k^\alpha$ exponentially and the TAMSD converges to $2\langle x^2\rangle_\textrm{th}$ with rate $\Delta^{\alpha-2}$ ($\nu_1=\beta=\alpha-2$), consistent to the results in Ref. \cite{JeonMetzler:2012}. While for SBM, we find $\langle x^2(t)\rangle\simeq \alpha Kk^{-1}t^{\alpha-1}$ depending on $t$ and $\langle \overline{\delta^2(\Delta)} \rangle\simeq 2Kk^{-1}T^{\alpha-1}$ being a $T$-dependent constant ($\gamma=\alpha-1$), which have been derived in Ref. \cite{JeonChechkinMetzler:2014}.

Note that the SBM with $0<\alpha<1$ is the mean-field approximation of the subdiffusive CTRW, as it describes the rescaled mean position of a cloud of walkers performing subdiffusive CTRW motion \cite{ThielSokolov:2014}. Therefore, the SBM shares the same correlation function \eqref{FBM} as the one of CTRW. Even for the great similarity between SBM and subdiffusive CTRW, they respond to the harmonic potential differently, since the EAMSD of the latter one tends to a thermal value and TAMSD grows as $\Delta^{1-\alpha}$. Differences among them come from how the external force acts on the system. The force does not affect on the particle at trap period in subdiffusive CTRW model. Otherwise, the MSDs will be the same as those of confined SBM, if the force acts on subdiffusive CTRW particles throughout the whole measurement time \cite{ChenWangDeng:2019-2}.

%\section{L\'{e}vy walk in harmonic potential}\label{two}

\emph{L\'{e}vy walk.}--- For standard L\'{e}vy walk, the particle runs with constant velocity $v_0$ and reselects its direction at a random time. The running times between consecutive unidirectional flights are independent and drawn from the same probability density function, usually the power law form $\varphi(t)=\frac{1}{\tau_0}\frac{\alpha}{(1+t/\tau_0)^{1+\alpha}}$.
Consider the overdamped Langevin equation of L\'{e}vy walk in a harmonic potential
\begin{equation}\label{OverdampModel}
\begin{split}
\dot{x}(t)=-k x(t)+v^{\textrm{LW}}(t),
\end{split}
\end{equation}
where $v^{\textrm{LW}}(t)$ denotes the velocity process of free L\'{e}vy walk. See Supplemental Material \cite{SM} for the derivations of \eqref{OverdampModel} from the Langevin picture of L\'{e}vy walk \cite{WangChenDeng:2019}. Note that the overdamped condition \cite{VanKampen:1992} usually holds in biological systems, such as the movements within a protein where the friction is usually large.

The key point is to evaluate the VCF of free L\'{e}vy walk $\langle v^{\textrm{LW}}(t)v^{\textrm{LW}}(t+\tau)\rangle=v_0^2p_0(t,t+\tau)$ \cite{FroembergBarkai:2013}.
Here $p_0(t,t+\tau)$ denotes the probability density function that no renewal happens within time interval $(t,t+\tau)$, and its double Laplace transform ($t\rightarrow s, \tau\rightarrow u$) is \cite{GodrecheLuck:2001}
\begin{equation}\label{P0}
\begin{split}
\hat{p}_0(s, u)=\frac{s(1-\hat{\varphi}(u))-u(1-\hat{\varphi}(s))}
{su(s-u)(1-\hat{\varphi}(s))}.
\end{split}
\end{equation}
The symbol $\hat{\varphi}(s)$ is the Laplace transform of $\varphi(t)$ and expanded as
\begin{equation}\label{WTPDF}
  \hat{\varphi}(s) \simeq 1+\sum_{j=1}^{\lfloor \alpha\rfloor}\frac{(-1)^j}{j!}\langle\tau^j\rangle s^j+(-1)^{\lfloor \alpha\rfloor+1}As^\alpha
\end{equation}
for small $s$, where $\lfloor \alpha\rfloor$ denotes the largest integer not larger than $\alpha$ and $\langle\tau^j\rangle$ the $j$-th moment of running time, $\langle\tau\rangle=\tau_0/(\alpha-1)$, and $A=|\Gamma(1-\alpha)|\tau_0^\alpha$.

It is hard to directly perform inverse transform on \eqref{P0}, which implies the VCF will not be as easy as the single-scale ones of FBM or SBM. In this case, to obtain the asymptotic forms of \eqref{P0} becomes a better choice. The results for large $t$ and $\tau$ have been discussed in many references studying free L\'{e}vy walk \cite{FroembergBarkai:2013,FroembergBarkai:2013-2,GodecMetzler:2013,WangChenDeng:2019}, and there is
\begin{equation}\label{largetau}
\begin{split}
p_0(t, t+\tau)
&\simeq
                     \left\{
    \begin{array}{ll}
       \frac{\sin (\pi \alpha)}{\pi}B\left(\frac{t}{t+\tau}; \alpha,1-\alpha\right), & 0<\alpha<1, \\[2pt]
      \tau_0^{\alpha-1} [\tau^{1-\alpha}-(t+\tau)^{1-\alpha}], & 1<\alpha<2, \\[4pt]
\end{array}
  \right.
\end{split}
\end{equation}
where $B(x; a, b) =\int_0^x dt t^{a-1}(1-t)^{b-1}$ is the incomplete Beta function. Note that $p_0(t,t)=1$ and it cannot be recovered by taking $\tau=0$ in \eqref{largetau} when $1<\alpha<2$. Even it seems right for $\alpha<1$ at $\tau=0$, the asymptotic form for small $\tau$ is not exact.
The asymptotic form of $p_0(t,t+\tau)$ for large $t$ but small $\tau$ can be obtained through the inverse Laplace transform of \eqref{P0} with small $s$ but large $u$ (see Supplemental Material \cite{SM}), that is
\begin{equation}\label{smalltau}
\begin{split}
p_0(t, t+\tau)
\simeq
\left\{
    \begin{array}{ll}
       1-\frac{\tau_0^{1-\alpha} |g(\tau)|}{\Gamma(2-\alpha)\Gamma(\alpha)}t^{\alpha-1}, & 0<\alpha<1, \\[4pt]
      1- |g(\tau)| -\tau_0^{\alpha-1}|g(\tau)| t^{1-\alpha}, & 1<\alpha<2,
      \end{array}
  \right.
\end{split}
\end{equation}
where $g(\tau)=1-\tau_0^{\alpha-1}(\tau_0+\tau)^{1-\alpha}$ satisfying $g(0)=0$ so that $p_0(t,t)=1$.
Next, we will study the EAMSD and TAMSD of confined L\'{e}vy walk $x(t)$ for different $\alpha$ by using the asymptotic form of VCF in \eqref{largetau} and \eqref{smalltau}.

\begin{figure}[tbhp]
\begin{minipage}{0.35\linewidth}
  \centerline{\includegraphics[scale=0.262]{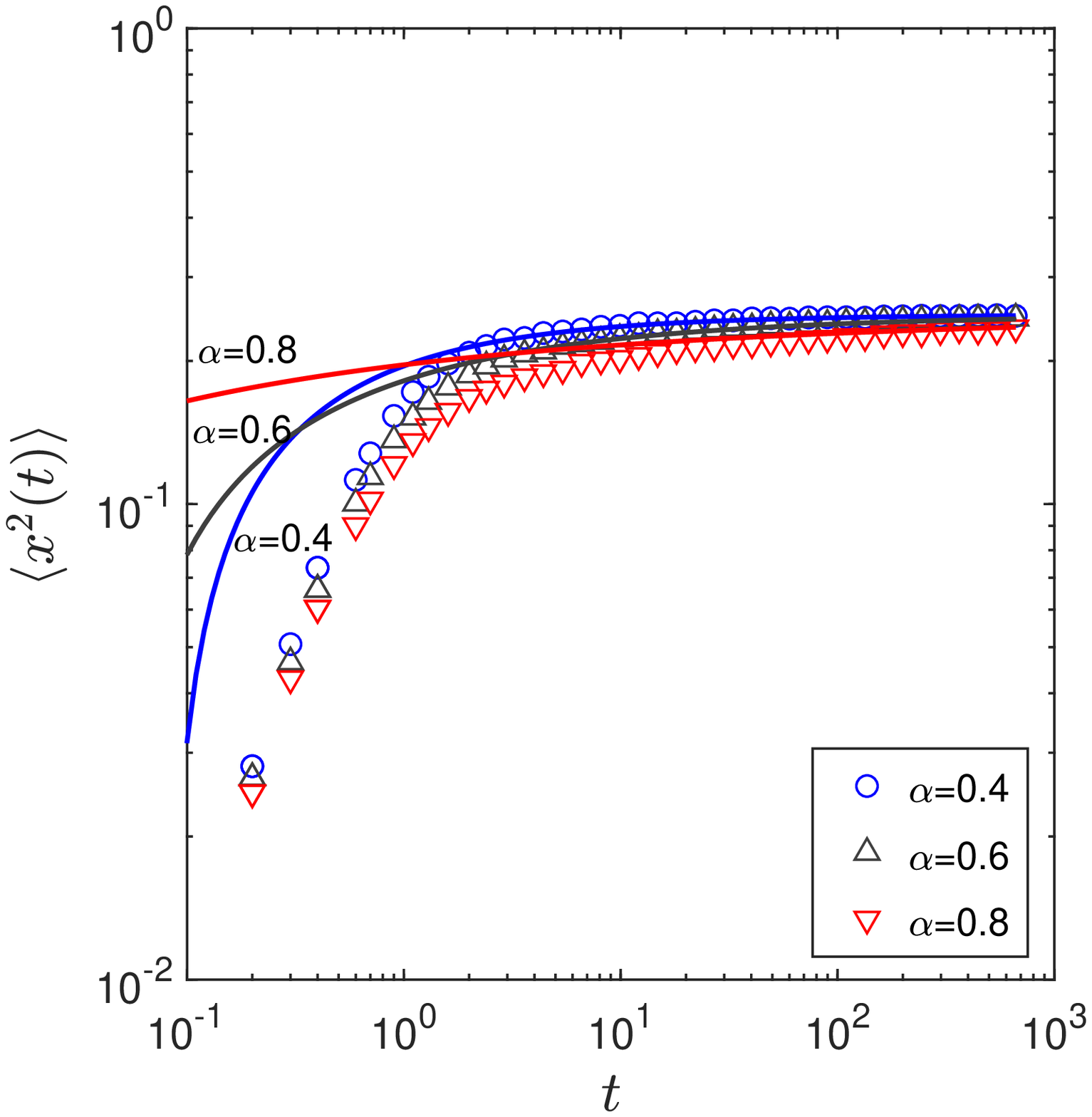}}
\end{minipage}
\hspace{1cm}
\begin{minipage}{0.35\linewidth}
  \centerline{\includegraphics[scale=0.262]{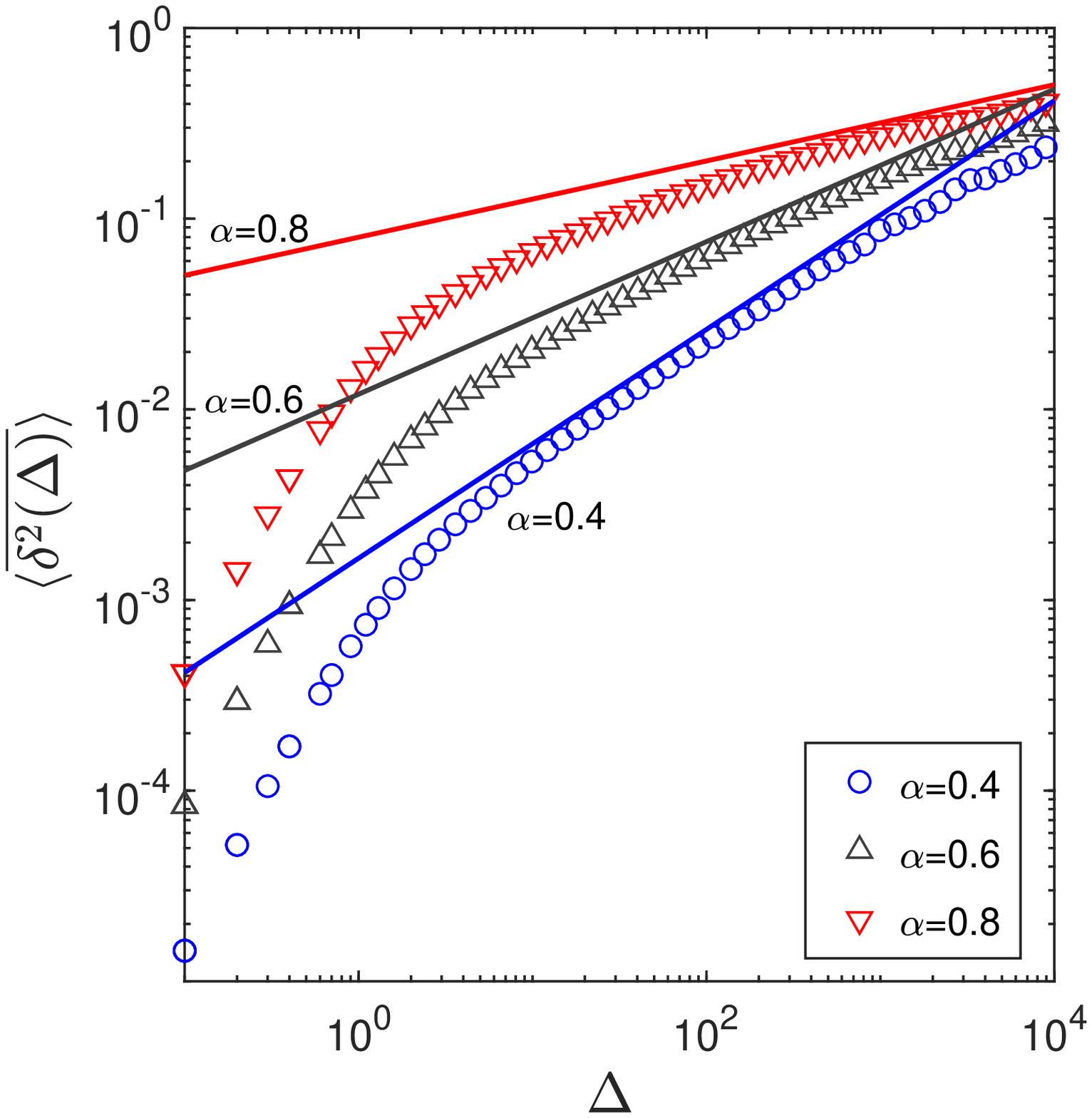}}
\end{minipage}
\caption{EAMSD and TAMSD of the confined L\'{e}vy walk for different $\alpha$. Color symbols represent the simulation results of EAMSD (left) and the ensemble-averaged TAMSD (right) with parameters $v_0=1$, $k=2$, and $\tau_0=0.1$. The color solid lines in two panels represent the asymptotic theoretical results in \eqref{xx01} and \eqref{TA01}, respectively. }\label{figure1}
\end{figure}

We firstly consider the ballistic L\'{e}vy walk with diverging time scale ($0<\alpha<1$). By comparing the VCF with small $\tau$ \eqref{smalltau} and the generic scaling form \eqref{EAMSD}, we obtain
\begin{equation}\label{xx01}
\begin{split}
\langle x^2(t)\rangle\simeq \frac{v_0^2}{k^2}-K_1t^{\alpha-1},
\end{split}
\end{equation}
where $K_1=v_0^2(e^{k \tau_0}k^{\alpha-1}\Gamma(2-\alpha,k \tau_0)-\tau_0^{1-\alpha})/(k^2 \Gamma(2-\alpha)\Gamma(\alpha))$ and $\Gamma(\beta,z)=\int_z^\infty dt e^{-t}t^{\beta-1}$ is the complementary incomplete Gamma function.
Obviously, EAMSD \eqref{xx01} grows to the stationary value $\langle x^2\rangle_{\textrm{th}}=v_0^2/k^2$ at power law rate $t^{\alpha-1}$.
The simulation results are shown in Fig. \ref{figure1}. The stationary value $\langle x^2\rangle_{\textrm{th}}$ is independent of $\alpha$. The speed of convergence becomes faster for a longer running time (a smaller $\alpha$). This feature resembles free L\'{e}vy walk very much, which displays ballistic diffusion $t^2$ with an $\alpha$-independent exponent.

Based on the result of EAMSD above, the ensemble-averaged TAMSD can be obtained by considering the large asymptotic form \eqref{largetau} additionally, i.e.,
\begin{equation}\label{TA01}
\begin{split}
\langle \overline{\delta^2(\Delta)}\rangle\simeq \frac{v_0^2}{k^2}\frac{2\sin (\pi \alpha)}{\alpha(1-\alpha)\pi}\left(\frac{\Delta}{T}\right)^{1-\alpha},
\end{split}
\end{equation}
for large $\Delta$ and $\Delta\ll T$. Instead of the saturation as EAMSD, it grows as $\Delta^{1-\alpha}$, which is similar to that of confined subdiffusive CTRW \cite{BurovMetzlerBarkai:2010,ChenWangDeng:2019-2}. This result is demonstrated in Fig. \ref{figure1}. It reveals some commonalities between ballistic L\'{e}vy walk and subdiffusive CTRW, though they have different asymptotic forms for small $\Delta$. When the lag time $\Delta$ approaches the measurement time $T$, this power-law growth stops and the function dips to the stationary value $\langle x^2\rangle_{\textrm{th}}$.

\begin{figure}[tbhp]
\begin{minipage}{0.35\linewidth}
  \centerline{\includegraphics[scale=0.262]{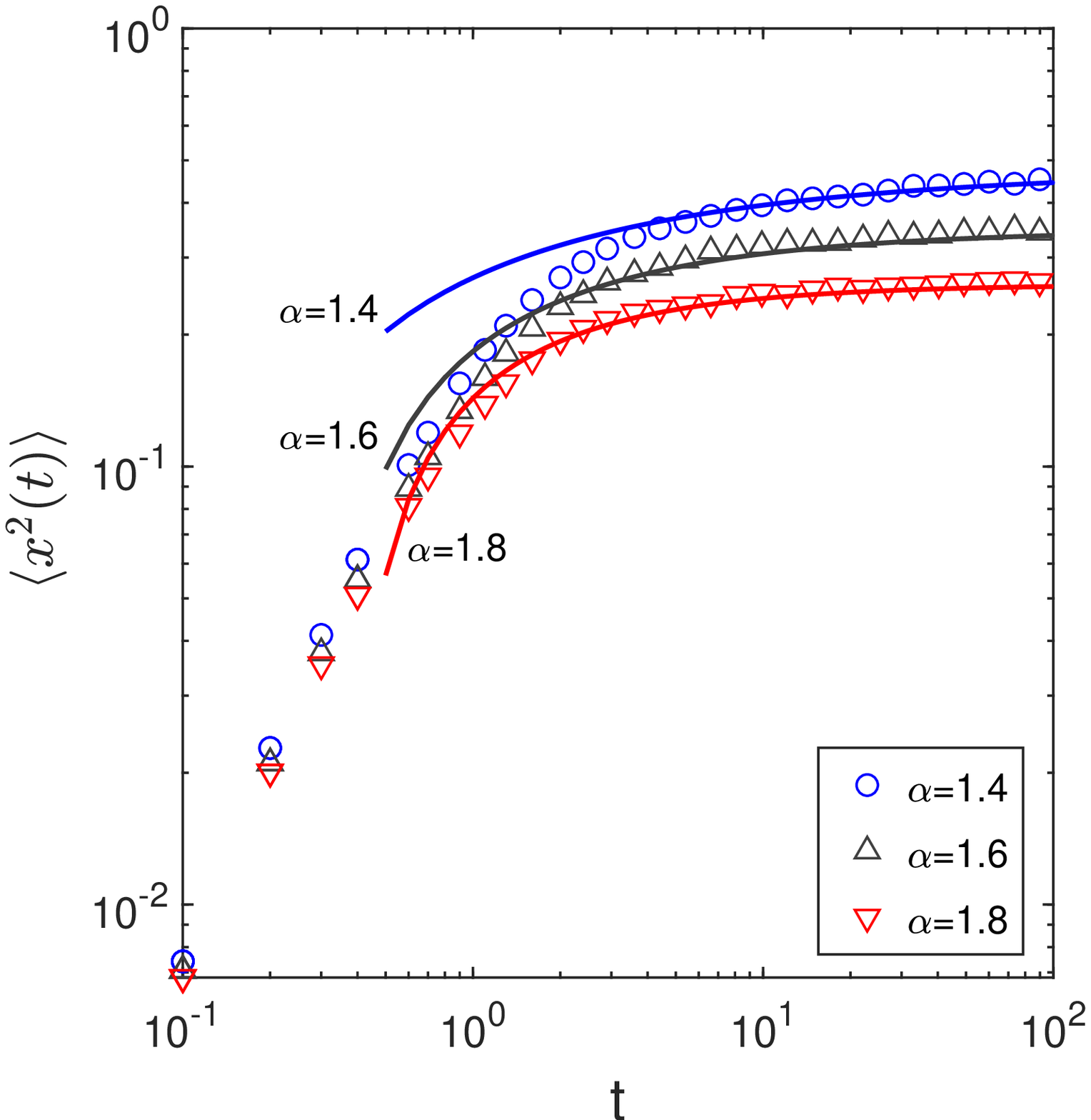}}
\end{minipage}
\hspace{1cm}
\begin{minipage}{0.35\linewidth}
  \centerline{\includegraphics[scale=0.262]{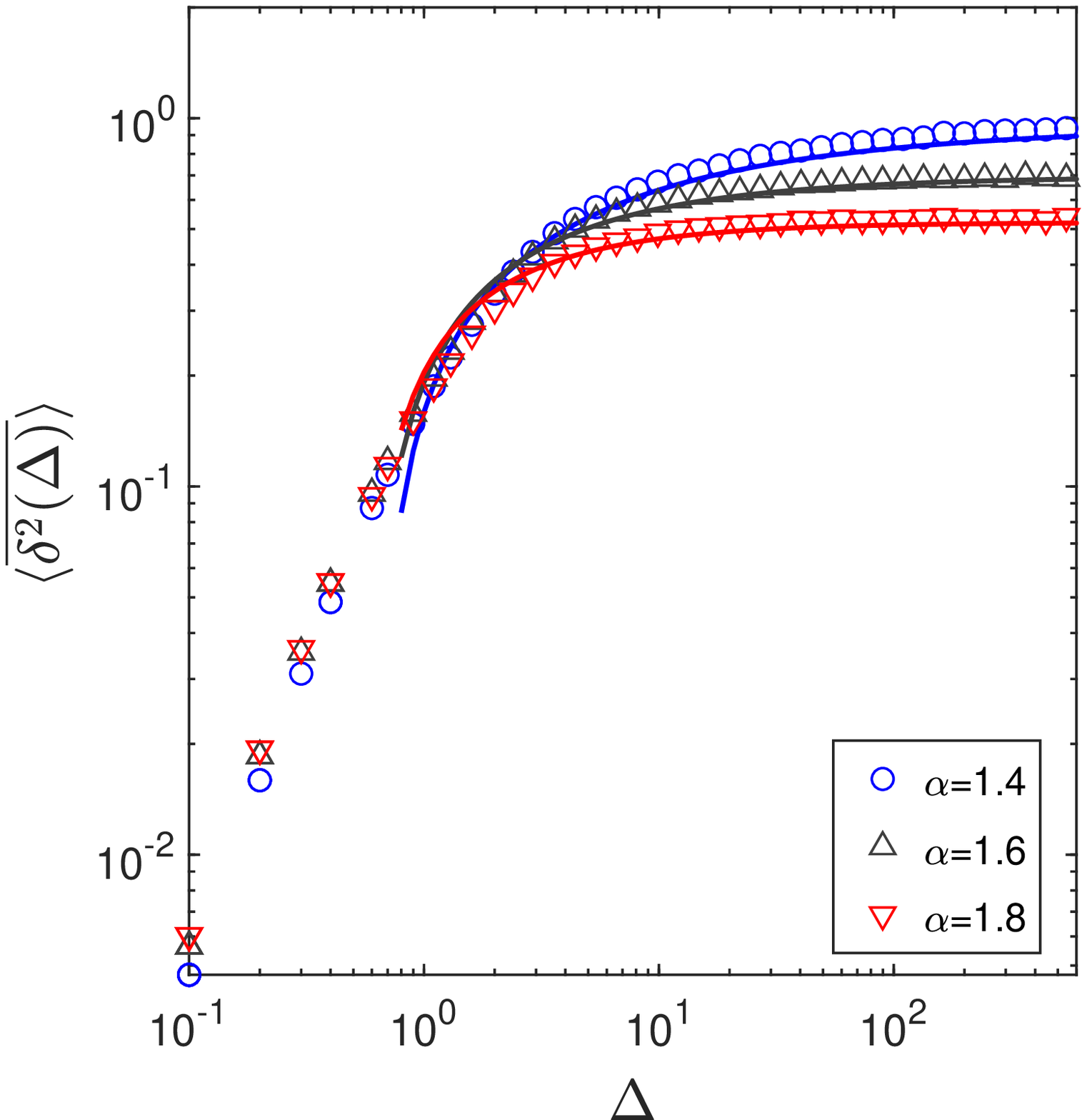}}
\end{minipage}
\caption{EAMSD and TAMSD of the confined L\'{e}vy walk for different $\alpha$. Color symbols represent the simulation results of EAMSD (left) and the ensemble-averaged TAMSD (right) with parameters $v_0=1$, $k=1$, and $\tau_0=0.1$. The color solid lines in two panels represent the asymptotic theoretical results in \eqref{xx12} and \eqref{asyTA12}, respectively. }\label{figure2}
\end{figure}

Enhanced L\'{e}vy walk with $1<\alpha<2$ is also an important process, which exhibits superdiffusion behavior. Its finite characteristic time scale endows it with some interesting phenomena, such as strong anomalous diffusion \cite{RebenshtokDenisovHanggiBarkai:2014}.
Similarly to the procedure for $\alpha<1$, now we obtain the EAMSD for large time $t$
\begin{equation}\label{xx12}
\begin{split}
\langle x^2(t)\rangle \simeq \langle x^2\rangle_{\textrm{th}}-K_2t^{1-\alpha},
\end{split}
\end{equation}
where $\langle x^2\rangle_{\textrm{th}}=v_0^2\tau_0^{\alpha-1}k^{\alpha-3}e^{k \tau_0}\Gamma(2-\alpha,k\tau_0)$ and $K_2=v_0^2(\tau_0^{\alpha-1}/k^2-\tau_0^{2\alpha-2}\Gamma(2-\alpha,\tau_0 k)e^{\tau_0 k}/k^{3-\alpha} )$.
The EAMSD \eqref{xx12} grows to the stationary value $\langle x^2\rangle_{\textrm{th}}$ at power law rate $t^{1-\alpha}$. Different from the case $\alpha<1$, now $\langle x^2\rangle_{\textrm{th}}$ decreases monotonously with respect to $\alpha$. And when $\alpha\rightarrow1$, it recovers the largest stationary value $\langle x^2\rangle_{\textrm{th}}=v_0^2/k^2$ in the case $\alpha<1$. The dependence of the thermal value on $\alpha$ reveals the correlation of free L\'{e}vy walk with different $\alpha$.
As Fig. \ref{figure2} shows, a larger $\alpha$ corresponds to a smaller stationary value and a faster relaxation. The EAMSDs with different $\alpha$ all increase to the stationary value slowly due to the algebraical relaxation.

As for the TAMSD, its ensemble average exhibits the same power law relaxation $\Delta^{1-\alpha}$ to the limiting value $2\langle x^2\rangle_{\textrm{th}}$ as the EAMSD in  \eqref{xx12}
\begin{equation}\label{asyTA12}
\begin{split}
\langle \overline{\delta^2(\Delta)} \rangle \simeq 2\langle x^2\rangle_{\textrm{th}}-\frac{2v_0^2\tau_0^{\alpha-1}}{k^2}\Delta^{1-\alpha}.
\end{split}
\end{equation}
In Fig. \ref{figure2}, we present the TAMSD for three different $\alpha$, which approaches the thermal value $2\langle x^2\rangle_{\textrm{th}}$ algebraically. The phenomenon that TAMSD is twice the EAMSD has been observed in the confined FBM and overdamped fractional Langevin equation \cite{JeonMetzler:2012}. The reason might be the finite characteristic time scale they both share. It is known that free FBM and fractional Langevin equation are ergodic \cite{DengBarkai:2009}. Similarly, the enhanced L\'{e}vy walk is also ergodic  if an equilibrium initial ensemble is set up \cite{GodecMetzler:2013}, since the ergodicity breaking parameter tends to zero as the measurement time $T\rightarrow\infty$. But if without a finite characteristic time scale, such as confined L\'{e}vy walk with $\alpha<1$, the TAMSD exhibits a power law growth, rather than converging to twice the thermal value.

\begin{figure}[tbhp]
\begin{minipage}{0.35\linewidth}
  \centerline{\includegraphics[scale=0.268]{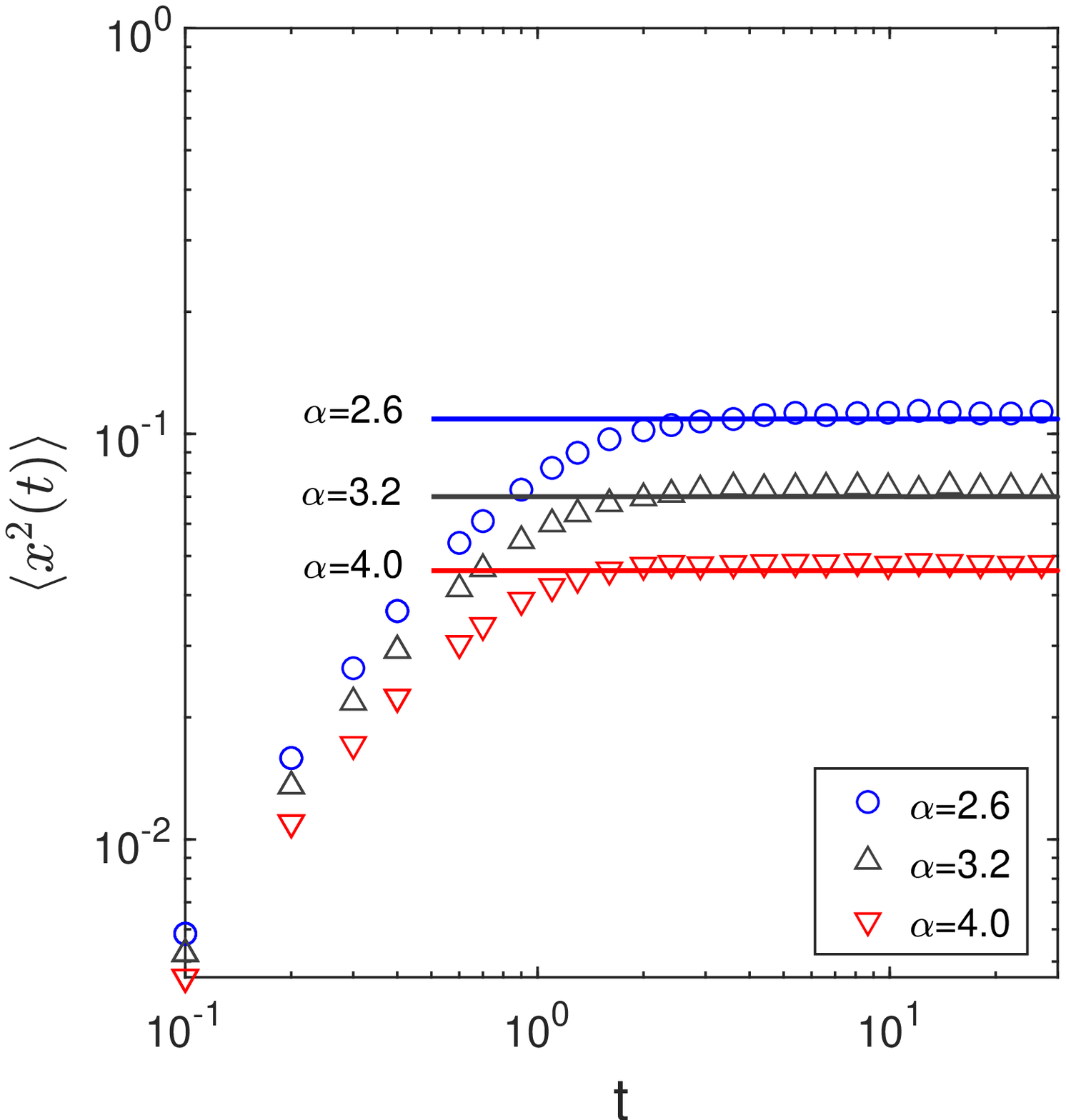}}
\end{minipage}
\hspace{1cm}
\begin{minipage}{0.35\linewidth}
  \centerline{\includegraphics[scale=0.268]{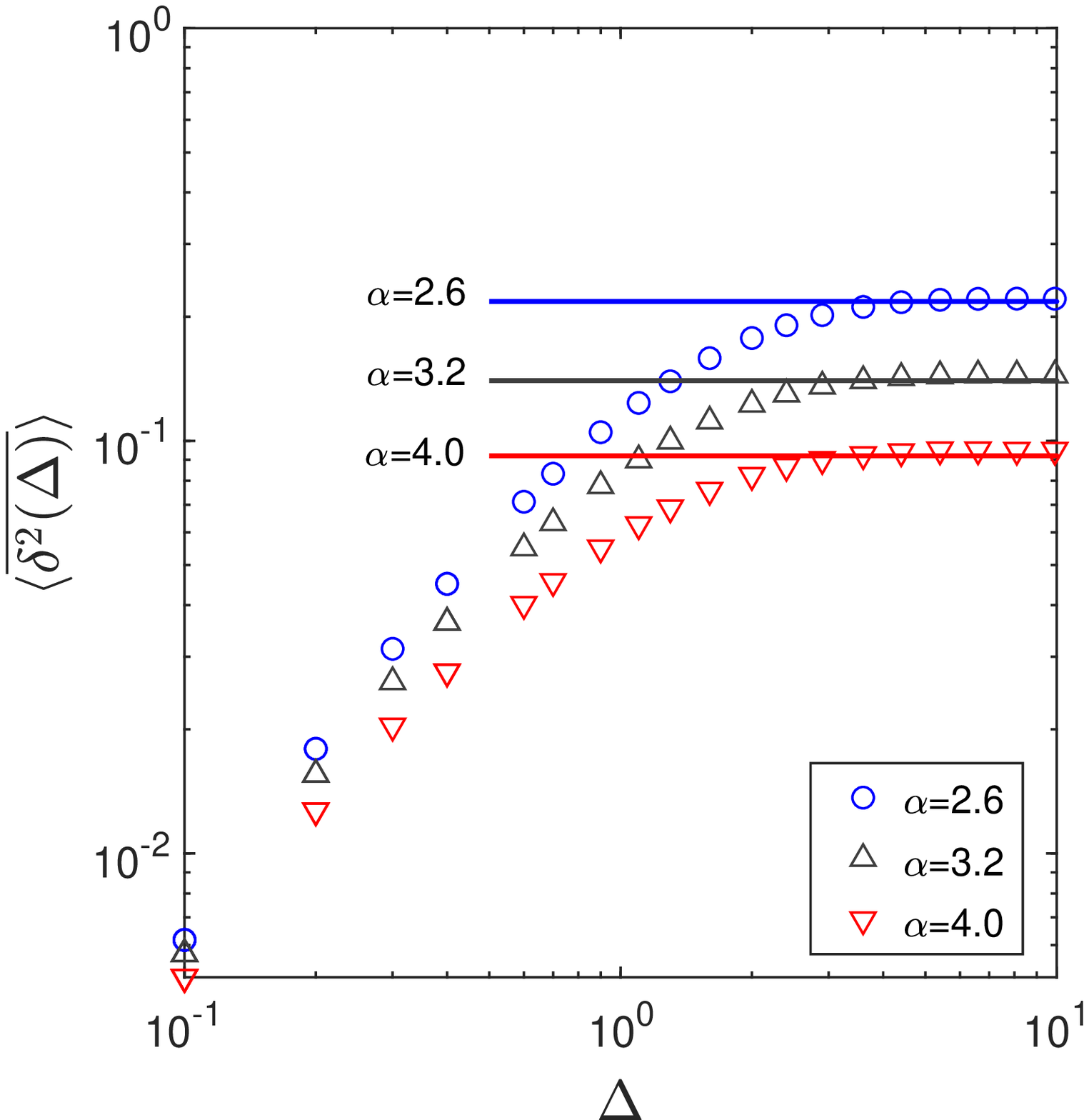}}
\end{minipage}
\caption{EAMSD and TAMSD of the confined L\'{e}vy walk for different $\alpha$. Color symbols represent the simulation results of EAMSD (left) and the ensemble-averaged TAMSD (right) with parameters $v_0=1$, $k=1$, and $\tau_0=0.1$. The color solid lines in two panels represent the stationary values in \eqref{xxbig2}. }\label{figure3}
\end{figure}

It is known that the free L\'{e}vy walk with $\alpha>2$ displays normal diffusion. Due to the finite characteristic time scale, similar results to $1<\alpha<2$ are expected, and there are
\begin{equation}\label{xxbig2}
\begin{split}
\langle x^2(t)\rangle \simeq \langle x^2\rangle_{\textrm{th}}, \quad  \langle \overline{\delta^2(\Delta)}\rangle \simeq  2\langle x^2\rangle_{\textrm{th}},
\end{split}
\end{equation}
for large time $t$ and $\Delta$, where the stationary value $\langle x^2\rangle_{\textrm{th}}$ is the same as the one for $1<\alpha<2$ in \eqref{xx12}. But the exponential relaxation dynamics for both EAMSD and TAMSD are observed in Fig. \ref{figure3} since VCF decays rapidly for $\alpha>2$, different from the power law relaxation for $1<\alpha<2$.

%\section{Summary}\label{six}

\emph{Summary.}---
We have built the relationship between the relaxation dynamics for a generic process confined in a harmonic potential with the VCF of the original process. The scaling forms of VCF for different range of lag time $\tau$ play some particular roles. The small $\tau$ behavior fully decides the EAMSD while the large $\tau$ behavior makes a nonnegligible contributions to TAMSD.
Our results are valid for both the processes with single-scale correlation function and the multi-scale ones, e.g., L\'{e}vy walk. Distinct relaxation dynamics have been detected for L\'{e}vy walk with different $\alpha$. %The scaled correlation functions observed in numerous systems become more valuable based on our results.
According to our observations/results, the scaled correlation functions in numerous systems become more valuable and some of their functions have been explicitly presented.

%\section*{Acknowledgments}
This work was supported by the National Natural Science Foundation of China under grant no. 11671182, and the Fundamental Research Funds for the Central Universities under grant no. lzujbky-2018-ot03.

%\appendix

%\section*{References}
\bibliographystyle{apsrev4-1}
\bibliography{ReferenceW}

\end{document}